\documentclass{llncs}
\usepackage{graphicx}                % allow us to embed graphics files
\graphicspath{{pictures/}} % where to search for the images

\PassOptionsToPackage{warn}{textcomp}  % to address font issues with \textrightarrow
\usepackage{textcomp}                  % use better special symbols
\usepackage{cite}                      % needed to automatically sort the references
\usepackage{tabu}                      % only used for the table example
\usepackage{booktabs}                  % only used for the table example
\usepackage{compress2}
\usepackage{url}

\usepackage{subfig}
\newcommand{\sflabel}[1]{(#1)}
\usepackage{amsmath,amssymb}
\usepackage{mathtools}
\usepackage{bm}
\usepackage{xspace}
\usepackage{xcolor}
\usepackage{datatool}
\usepackage{siunitx,textcomp}
\newcommand{\numberset}{\mathbb}

\newcommand{\Z}{\numberset{Z}}
\newcommand{\R}{\numberset{R}}
\newcommand{\vect}[1]{\bm{#1}}
\newcommand{\ea}{\textit{et al.}\xspace}

\newcommand{\visone}{\texttt{visone}\xspace}

\begin{document}

%% Paper title.
\title{Drawing Dynamic Graphs Without Timeslices}

%% This is how authors are specified in the journal style

%% indicate IEEE Member or Student Member in form indicated below
\author{Paolo Simonetto\inst{1}
\and Daniel Archambault\inst{1}
\and Stephen Kobourov\inst{2}
}
%\authorfooter{
%% insert punctuation at end of each item
\institute{Swansea University, UK\\ \email{paolo.simonetto@swansea.ac.uk,d.w.archambault@swansea.ac.uk}
\and University of Arizona, US\\ \email{kobourov@cs.arizona.edu}
}
\authorrunning{P. Simonetto {\it et al.}}
\titlerunning{Drawing Dynamic Graphs Without Timeslices}

\maketitle

%% Abstract section.
\begin{abstract}
  Timeslices are often used to draw and visualize dynamic graphs.  While timeslices are a natural way to think about dynamic graphs, they are routinely imposed on continuous data. Often, it is unclear how many timeslices to select: too few timeslices can miss temporal features such as causality or even graph structure while too many timeslices slows the drawing computation. We present a model for dynamic graphs which is not based on timeslices, and a dynamic graph drawing algorithm, DynNoSlice, to draw graphs in this model. 
  %Nodes and edges, along with their attributes, can be defined to evolve in continuous intervals of time. 
  In our evaluation, we demonstrate the advantages of this approach over timeslicing on continuous data sets.
\end{abstract}

%%%%%%%%%%%%%%%%%%%%%%%%%%%%%%%%%%%%%%%%%%%%%%%%%%%%%%%%%%%%%%%%
%%%%%%%%%%%%%%%%%%%%%% START OF THE PAPER %%%%%%%%%%%%%%%%%%%%%%
%%%%%%%%%%%%%%%%%%%%%%%%%%%%%%%%%%%%%%%%%%%%%%%%%%%%%%%%%%%%%%%%

%On the left, a large number of slices is used for higher accuracy. Unfortunately, this greatly increase the input size and data redundancy, since many timeslices only present change in a few elements. On the right, data is flattened into few timeslices to reduce input size. Much information is lost, since from the two slices we cannot deduce that $b$, $c$, $d$, and $e$ never appear at the same time or in this order. 

%% Keywords that describe your work. Will show as 'Index Terms' in journal
%% please capitalize first letter and insert punctuation after last keyword
%\keywords{Dynamic graph drawing, continuous time, graph visualization}

%% ACM Computing Classification System (CCS). 
%% See <http://www.acm.org/class/1998/> for details.
%% The ``\CCScat'' command takes four arguments.

%\CCScatlist{ % not used in journal version
% \CCScat{K.6.1}{Management of Computing and Information Systems}%
%{Project and People Management}{Life Cycle};
% \CCScat{K.7.m}{The Computing Profession}{Miscellaneous}{Ethics}
%}

%% Uncomment below to include a teaser figure.

\section{Introduction}

%Graphs are mathematical objects used to represent relational data. They are composed of a set of elements (nodes) and pair-wise relations (edges) between them.  Graphs are typically represented graphically with node-link diagrams, where the nodes are glyphs (circles) and edges are curves (segments) connecting them. In such drawings, the position of the nodes can significantly influence the readability of the underlying data. This is one the main motivations for studying how to place the nodes and route the edges in order to make the underlying patterns and trends in the data easier to see. 
% order to promote desired characteristics in the final drawing.  

Graphs offer a natural way to represent a static set of relations. In order to encode changes to a graph over time, %(adding/removing nodes or edges), 
static graphs can be extended to dynamic graphs. Dynamic graphs are traditionally thought of as a sequence of static graphs in a finite number of moments in time, or \emph{timeslices}. We refer to these dynamic graphs as \emph{discrete dynamic graphs}.
Discrete dynamic graphs are widely used for several reasons. Drawing timesliced graphs is similar to drawing static graphs, allowing force-directed approaches to be easily adapted to dynamic data.  Also, timeslicing works fine for data that changes at discrete, regular time intervals.

%Discrete dynamic graphs, however, are not suited to represent data that changes in a non-discrete fashion. When data changes occur outside a small number of moments in time, we need to increase the slicing frequency or to flatten the data into the closest timeslice (see Figure~\ref{FIG:Teaser}).  Neither approach is ideal. In the first case, we might be required to drastically increase the number of slices, beyond the typical values for discrete dynamic graphs (10 to 30 timeslices). The oversampled data is mostly redundant as a timeslice might be introduced to describe the change of very few elements. In the second case, we might lose important information when flattening an interval to a single timeslice. For example, by aggregating all the changes in a month of data in a single timeslice, we will lose any information about event causality or propagation (since all changes will happen at once) or about recurring patterns (nodes that appear and disappear every week).
%
Algorithms to draw discrete dynamic graphs strike a balance between graph drawing readability and stability~\cite {archambault2016can}. Readability requires the drawing of individual timeslices to be of high quality while stability (mental map preservation) requires nodes to not move too much between consecutive timeslices % positions should be stable among consecutive timeslices 
for easy identification~\cite {12ArchambaultGD}. These requirements conflict with each other as node movement is required for timeslice readability, while it negatively impacts stability.
%is required for mental map preservation.  

%The approaches proposed for discrete dynamic graph drawing generally extend static graph drawing algorithms to take into account the preservation of the mental map. Typically, each timeslice is a static graph that is individually drawn in combination with the timeslices directly before and after with the goal of stabilizing node positions. The % large and 
%successful use of such techniques suggests that they are adequate for representing dynamic graphs that change in a discrete fashion.
%
%When the data set is continuous, however, two new considerations must be taken into account. First, a number of slices must be selected carefully based on the data (see Figure~\ref{FIG:Discretisation_issues}) which is often non-obvious.  Second, even if a large number of timeslices is chosen, we are generally still required to perform data discretization when building the timeslices. This operation implicitly aggregates data along the time dimension, which necessarily hides part of the original information. For example, slices created by summing values over year quarters will necessarily lose the information regarding the individual months.

\begin{figure}[t]
  \begin{center}
\subfloat[Discrete Dynamic Graph]{\includegraphics[width=.43\textwidth]{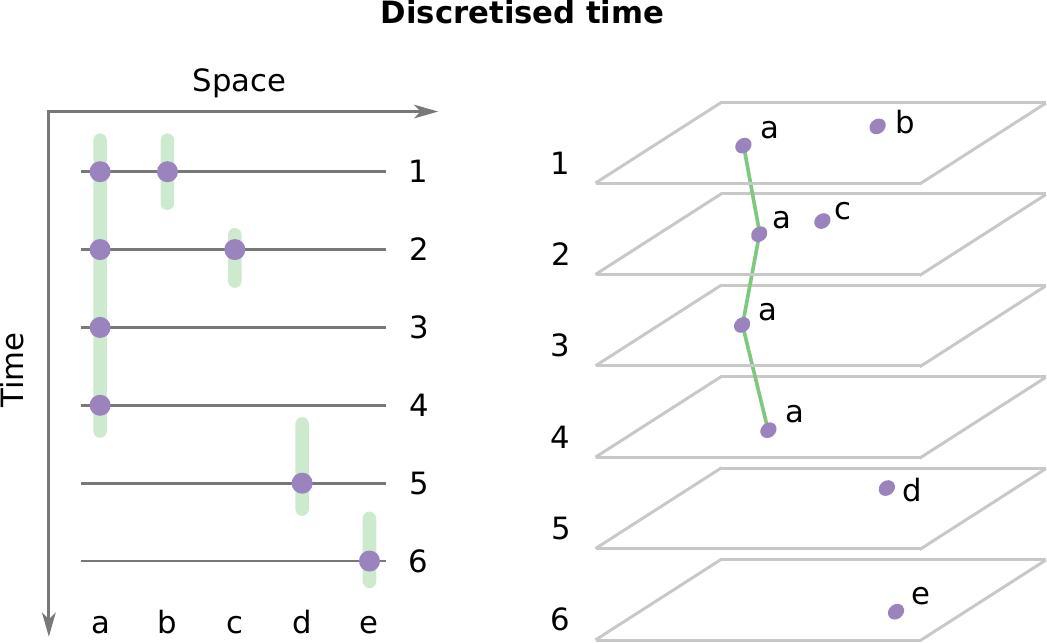}} \hfill
\subfloat[Continuous Dynamic Graph]{\includegraphics[width=.53\textwidth]{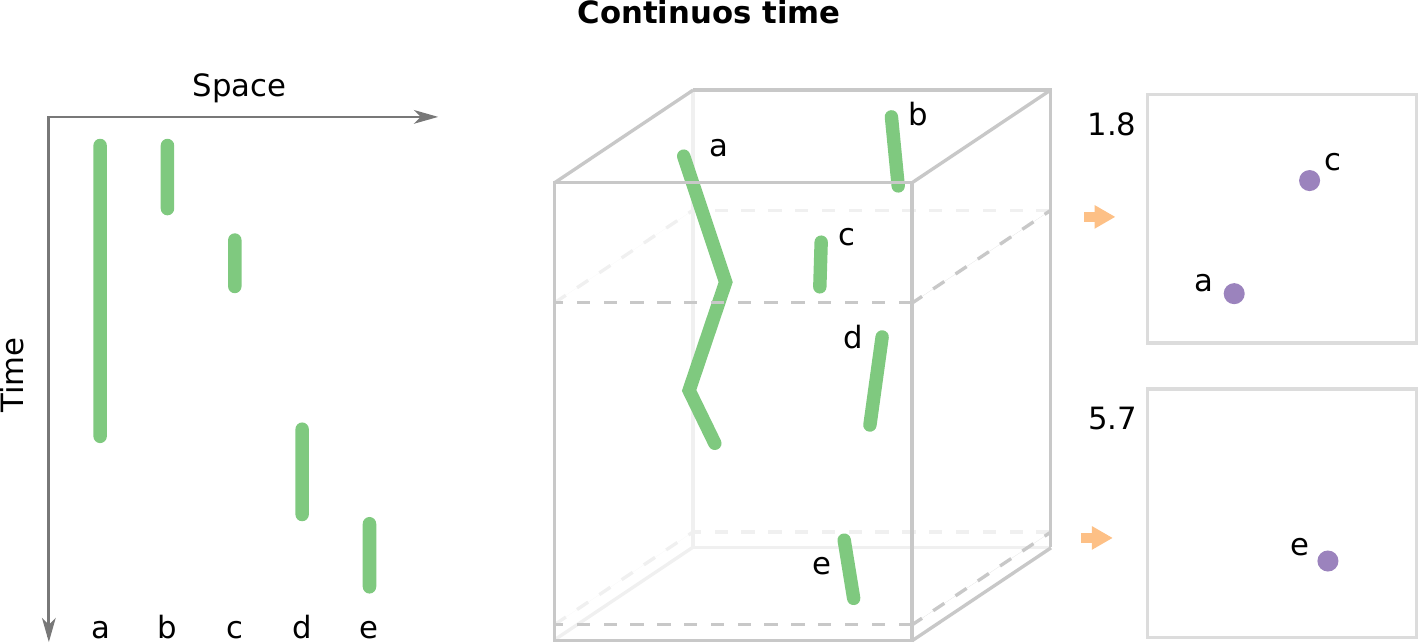}}
\medskip\caption{\small %Drawing dynamic graphs without timeslices. Let us 
  A dynamic graph with 5 nodes. % ($a$--$e$).  In all figures nodes are green bars. 
  (a) A discrete representation with nodes projected onto timeslices.  Projecting the nodes (purple dots) results in information loss due to aggregation across time.
  (b) In our approach, nodes are defined as piecewise linear curves in the space-time cube, as timeslices are not imposed on continuous data.} 
\label{FIG:Teaser}
  \end{center}
\end{figure}

Approaches for discrete dynamic graph drawing extend static graph drawing algorithms.  Each timeslice is a static graph that is drawn considering the timeslices before and after to stabilize the transition.  However, when the time dimension is continuous, there are no timeslices.  Thus, evenly spaced timeslices are selected and the graph elements are projected to the closest one.   This operation implicitly aggregates data along the time dimension (Fig.~\ref{FIG:Teaser}).  Selecting too many timeslices leads to slow layout computation while selecting too few leads to information loss due to projection errors when events are of short duration, obscuring graph structure.  %However, 
Even for graphs with nodes and edges that span long periods of time, regular timeslices can obscure important temporal patterns.  %It is also impossible to generate drawings for timeslices not present in this aggregation.  For example, if we would like to view the data at time $1.5$, we would need to generate a new discrete dynamic graph from the continuous data with a larger number of timeslices.

A natural solution to the problems above would be to refrain from imposing timeslices on the continuous data and draw the graph directly along a continuous time dimension.  In a recent state-of-the-art report, Beck \ea~\cite[p. 15]{STAR_DyGraph_visualisation} states \emph{``the effects of using continuous time with arbitrary fine sampling rates, rather than discretized time, are largely unexplored''}. %In this paper,
In this paper, we introduce a model for dynamic graph drawing that does not use timeslices.  We call these graphs \emph{continuous dynamic graphs} and propose the first dynamic graph drawing algorithm, DynNoSlice, to draw them along a continuous time dimension.
%directly in continuous time.  
%As a result, the graph is drawn once and timeslices can be selected at any point in time.  %Although a natural way to visualize continuous graphs is an animation, we also develop a method for selecting interesting timeslices in order to support a small multiples visualization of these data sets.
%Our contribution is to create this model for dynamic graph drawing without timeslices. We call these graphs \emph{continuous dynamic graphs} and propose a dynamic graph drawing algorithm, DynNoSlice, to draw them directly in continuous time.  Instead of imposing timeslices at fixed intervals, this new model allows for nodes and edges to adapt, %data set complexity to change through time 
%depending on how individual nodes interact in the data set.  The approach operates on a dynamic graph with a continuous time dimension, where nodes are piecewise linear functions in time that can be drawn without timeslices (see Fig.~\ref{FIG:Teaser_2}).  As a result, the graph is drawn once and timeslices can be selected at any time to produce representations of that data at that time.  Although a natural way to visualize continuous graphs is an animation, we also develop a method for selecting interesting timeslices in order to support a small multiples visualization of these data sets.  In our evaluation, we demonstrate that our continuous approach has significant advantages over timeslicing the continuous data.

\section{Related Work}
Dynamic graphs visualization is a well established field of research, as shown in the recent survey by Beck~\ea~\cite{STAR_DyGraph_visualisation}.  If we were to place our approach in this model, it would be an ``offline time-to-time mapping'' without timeslices, as it works with full knowledge of the input data. Related approaches also include the ``superimposed time-to-space mapping'', as this approach inherently defines a space-time cube. 
%Here we will present some of the more relevant papers of this type.

{\bf Offline Drawing.} Offline dynamic graph drawing algorithms capture all of the dynamic data beforehand and optimize across it simultaneously.  Foresighted layout algorithms~\cite{Foresighted_dyGraph_layout,Forsighted_dyGraph_layout_with_tolerance} are among the first attempts at offline dynamic graph drawing.  They create a supergraph as the union of the elements at each timeslice. The supergraph is then used to define the position of nodes and edge bends. Mental map preservation is the primary focus of this approach.
%, but might not offer sufficient flexibility across many timeslices in the data.

Brandes and Corman~\cite{Unrolling_of_network_evolution} designed an approach for offline dynamic graph drawing that visually refers to time as a third, discrete spacial component. %In their methods, 
Timeslices are placed on top of each other, %as if drawn on transparent slides, 
forming a layered space-time cube. 
%Nodes keep their position across timeslices and are visualized as cylinders that intersect the timeslices in which they appear. 
%The space-time cube can be rotated and the slides can be removed to study the network evolution. 
A similar technique is also used by Dwyer and Eades~\cite{Columns_and_worms}.  In both approaches, the time dimension is modeled using timeslices.

Erten~\ea~\cite{Simultaneous_graph_drawing} explored several ways to superimpose different graphs, including the use of different colors for different timeslices of a layered space-time cube. In GraphAEL~\cite{GraphAEL,GraphAEL_system}, the space-time cube was created by connecting the same nodes in consecutive timeslices with inter-timeslice edges and optimizing this structure with a force-directed algorithm. A similar technique is used by Groh~\ea~\cite{Dyson} and by Itoh~\ea~\cite{Bloggers_changes} for social networks. Brandes and Mader~\cite{Stress_comparison} perform a metric-based evaluation of several strategies for drawing discrete dynamic graphs including aggregation, anchoring, and linking.  Linking strategies performed best in balancing drawing quality (measured using stress) and mental map preservation (measured using distance traveled).
%Even methods that do not explicitly use this metaphor as a basis for drawing or visualization are closely related to timeslicing.
Rauber~\ea~\cite{Dynamic_tSNE}, identify vectors $v_i[t]$ in the cost gradient whose geometrical interpretation would be identical when working in a layered space-time cube.

Our approach can be seen as an extension of offline discrete dynamic graph drawing approaches to continuous time. However, the difference is substantial as in related work timeslices are imposed on the data.  By drawing in continuous time, each node and edge defines its own trajectory with its own complexity whereas discrete algorithms impose linear interpolations between timeslices.

%The resulting 3D layout is then optimized with a force-directed algorithm to improve mental map preservation. Our drawing algorithm can be seen as an extension of GraphAEL to continuous time. However, the difference is quite substantial, as continuous time allows a more fine definition of node trajectories, whereas GraphAEL only allows for a linear interpolation between timeslices.
%
%A layered space-time cube is also employed by Groh~\ea~\cite{Dyson} and by Itoh~\ea~\cite{Bloggers_changes} to show blogging activity. Even other methods that do not explicitly use this metaphor as a basis for drawing or visualization techniques are closely related to timeslicing. For example, the adaptation of t-SNE to dynamic graph drawing by Rauber~\ea~\cite{Dynamic_tSNE}, identifies vectors $v_i[t]$ in the cost gradient whose geometrical interpretation would be identical when working in a layered space-time cube rather than on a sequence of frames.

%Brandes and Mader~\cite{Stress_comparison} perform a metrics-based evaluation of several strategies for drawing dynamic graphs including aggregation, anchoring, and linking.  In their evaluation, they consider an offline scenario and draw all test data sets using the same stress minimization approach.  Linking strategies performed best in balancing drawing quality (measured using stress) and mental map preservation (measured using distance traveled). % and the results indicated that linking strategies performed well.  
%In this paper, we perform similar metric-based evaluation, adapted to our continuous dynamic graph model.

Time-to-space mapping approaches~\cite{STAR_DyGraph_visualisation} use one dimension for space and one dimension for time in order to visualize the network~\cite{Semantic_substrates,Parallel_edge_splatting,Massive_sequence_views,EgoNetCloud}. Recent scalable approaches use dimensionality reduction to show time as a curve in the plane~\cite {15Bach,15Elzen}. By using spatial dimensions to represent time, these visualizations are substantially different from classic dynamic node-link diagrams.
%By drawing the network along a single dimension plus time, these methods are forced to either work with very small graphs, or adopt drawing solutions that are substantially different from classical node-link diagrams.

%\subsubsection{Online Drawing}
%Another related category of algorithms is online drawing dynamic graph drawing. 
{\bf Online Drawing.} Given an incoming stream of data, online drawing algorithms continuously update the current graph drawing to take into account new data. 
%The problem has been initially been addressed by In an early approach, 
Misue \ea~\cite{Layout_adjustment} optimize mental map preservation by maintaining the horizontal and vertical order of the nodes through adjustments. Frishman and Tal~\cite{Online_Frishman} proposed a different force-directed algorithm which could also be partially executed on the GPU. Gorochowski \ea~\cite{Online_Aging} use node aging to decide how much a node position should be preserved at a given time. Finally, Crnovrsanin \ea~\cite{Online_Multilevel} adapted the \texttt{FM\textsuperscript{3}}~\cite{FM3} multilevel approach to an online setting.

Although related, online approaches are %substantially 
different as they operate under more stringent constraints due to lack of access to future information. %such as slices at time greater than the current one. 
%whereas in the offline setting the entire graph evolution is known in advance.
%Second, online approaches have much tighter time constraints, as they are often expected to operate in real time.
%Although our approach might seem similar to the online graph drawing algorithms, the data used by our approach is different.  Online algorithms can only consider temporal information in the current timeslice or in the past.  
Offline algorithms, such as DynNoSlice, %take full advantage of 
use the full knowledge of the graph evolution.
%is known in advance.consider all information across a time interval simultaneously.
%allowing the influence of older data to be diminished.

%\subsection{3D Graph Drawing}
{\bf 3D Graph Drawing.} Several 2D algorithms have been extended to 3D. Bru\ss{} and Frick~\cite{Gem3D} proposed \texttt{Gem3D}, which extends \texttt{Gem}~\cite{Gem} to 3D. Cruz and Twarog~\cite{3D_simulated_annealing} extended the simulated annealing approach of Davidson and Harel~\cite{2d_simulated_annealing} to 3D. Other algorithms  %naturally 
work in both 2D and 3D, such as \texttt{GRIP} by Gajer and Kobourov~\cite{GRIP}. 
%Some algorithms are specifically designed for 3D. In particular, 
Munzner~\cite{Hyperbolic_3D} proposed an algorithm to draw directed graphs in a 3D hyperbolic space. Several other algorithms deal with particular constraints, such as orthogonal~\cite{3D_orthogonal}, or nested~\cite{3D_nested} drawings.  Cordeil \ea~\cite {16Cordeil} investigate the visualization of graphs in 3D using immersive environments.

Although our approach does compute a 3D layout of the dynamic graph, we do not produce a 3D visualization. % of the dynamic graph.  
Since our third dimension is time, 
%in our case and 
nodes and edges have additional constraints and are not free to move arbitrarily in 3D.
%three dimensional space.
%Thus, once again, although related, 3D approaches are substantially different from our approach because of the different nature of the underlying problems. 
%reason, the problems faced by the above algorithms are not the same as those considered by our method. 

%\subsection{Space-Time Cube Visualization and Manipulation}
%A few approaches, such as
{\bf Space-Time Cubes.}  Sallaberry \ea~\cite{Dynamic_graphs_visualisation} visualize dynamic graphs by manipulating the space-time cube. Several other information visualization methods %related or not to dynamic graphs, 
perform similar operations in the space-time cube, as discussed in Bach~\ea~\cite{16BachCGF}. %review most dynamic data publications, identifying operations and approaches used to also considers approached that operate on the space-time cube. 
These approaches assume that the space-time cube is already given and focus on how to accommodate it in 2D,  
%In the report, the authors assume that the space-time cube is already given and focus on how to accommodate it in 2D media. 
whereas our approach constructs the space-time cube for dynamic graphs using a continuous time dimension.
%focus on the construction of the space-time cube for dynamic graphs using continuous time, the paper of Bach~\ea~\cite{16BachCGF} %perfectly complements our work by providing additional insights into the analysis and visualization of the output from our approach.

\begin{figure}[t]
  \centering
\subfloat[]{\includegraphics[width=0.18\linewidth]{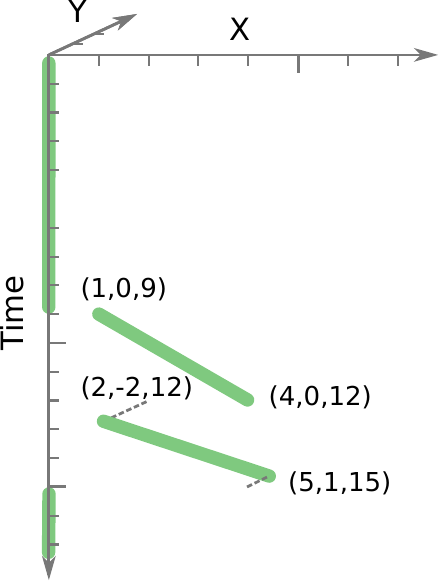}\label{FIG:Transformation_1}} \hspace{0.5cm} 
\subfloat[]{\includegraphics[width=0.18\linewidth]{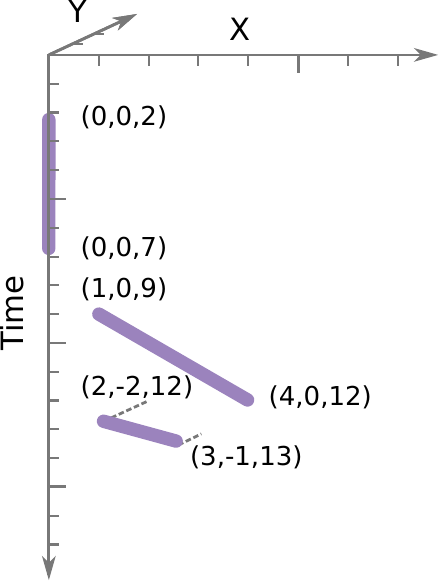}\label{FIG:Transformation_2}} \hspace{0.5cm} 
\subfloat[]{\includegraphics[width=0.18\linewidth]{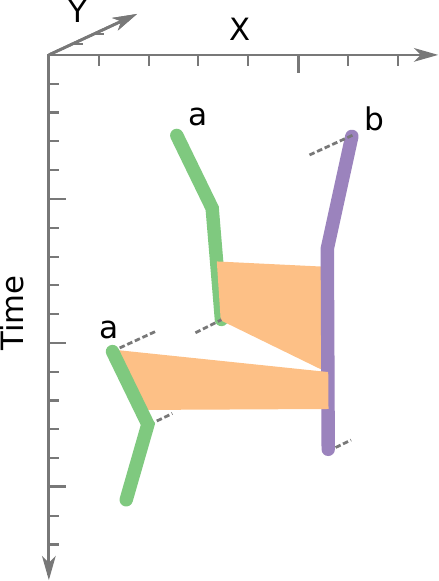}\label{FIG:Transformation_3}}
\caption{\small A continuous dynamic graph in the space-time cube. (a) Position attribute for a node $v$. %Piecewise linear functions encode node position across time. 
(b) The appearance attribute.  
%Node $v$ only appears over some intervals of time.  
(c) An edge in a continuous dynamic graph, % Edges connect node trajectories 
as encoded by the edge appearance attribute. 
%The edge traces one or more surfaces in the space-time cube.
} \label{FIG:Transformation}
\end{figure}

\section{Continuous Dynamic Graph Model}
Let $G=(V,E)$ be a static graph defined with node set $V$ and edge set 
%on a set of nodes $V$ and a set of 
%(directed or undirected) edges 
$E \subseteq V \times V$. We define \emph{attributes} (functions) on the nodes and edges of the graph that encode characteristics such as their positions, weights, and labels. The node position attribute ($\mathcal{P}_G : V \rightarrow \R^2$) maps a node $v$ into its position in the 2D plane $\vect{p}_v$, %\ie,  
e.g.,  $\mathcal{P}_G(v) = (1,4)$.  This attribute is not integral to our model of continuous dynamic graphs, but is used to compute and store the layout of these graphs.

We define a continuous dynamic graph $D = (V,E)$ as a graph whose attributes are also a function of time. Let $T$ be the time domain defined as an interval in $\R$. Attributes are functions defined in the domain $V \times T$ for nodes, and $E \times T$ for edges. For example, the node position attribute $\mathcal{P}_D : V \times T \rightarrow \R^2$ is the function that describes the position of $v$ for each time $t \in T$. We assume (w.l.o.g.) that the node and edge attributes can be defined piecewise, e.g.:
\[ \mathcal{P}_D(v,t) = \mathcal{P}_v(t) =  \begin{cases} \mathcal{P}_{v,1}(t) & \text{for } t \in T_1 \\ %\mathcal{P}_{v,2}(t) & \text{for } t \in T_2 \\\ 
\vdots \\ \mathcal{P}_{v,n}(t) & \text{for } t \in T_n \\ \vect{p}_{v,\omega} & \text{otherwise} \end{cases} \]

In other words, a dynamic attribute can be thought of as a map that links each node (edge) to a sequence of functions $\mathcal{P}_{v,i}$ that describe its behavior in disjoint intervals of time $T_i$, with a default value returned for $t \notin \bigcup_{i} T_i$. 
%While the model %graph definition 
%allows for the definition of arbitrary %functions, our approach only supports 
In the rest of paper we consider only piecewise linear functions for nodes and edges. 
%will be ruled surfaces.   For this reason, from now on, 
%In the rest of the paper, these functions are assumed to be piecewise linear.

Attributes specify %define 
a variety of node and edge characteristics. For data that can be meaningfully interpolated (e.g., colors, weights, positions), 
%each of the %case
the functions above can be described by initial and final values.
%, since each intermediate point can be computed by interpolation. % in the related interval. 
 For attributes without meaningful interpolation (e.g., labels), we prefer functions that are constant in the related interval. Position and label attributes for a node $v$ of $D$ can be:
\[ \mathcal{P}_v(t) = \begin{cases} (1,0) \rightarrow (4,0) & \text{for } t \in (9,12] \\ (2,-2) \rightarrow (5,1) & \text{for } t \in (12,15] \\ (5,1) \rightarrow (4,5) & \text{for } t \in [17,19] \\ (0,0) & \text{otherwise} \end{cases} 
\mathcal{L}_v(t) = \begin{cases} \text{Jane Doe} & \text{for } t \in (10,11] \\ \text{Jane Smith} & \text{for } t \in (11,16] 
\\ \text{unknown} & \text{otherwise} \end{cases} \]  

We define the attribute \emph{appearance} that implements the classic dynamic graph operations node/edge insertion and deletion. %Appearance is a Boolean attribute indicating the intervals in which a node $v$ (edge) is present in the graph, for example:
\[ \mathcal{A}_v(t) = \begin{cases} \text{true} & \text{for } t \in [2,7) \\ \text{true} & \text{for } t \in (9,13] 
\\ \text{false} & \text{otherwise} \end{cases} \]
In order to mark an edge $e=(u,v)$ as present in $T_x$, we need to ensure that both $u$ and $v$ are present for the entire $T_x$.

This definition supports changes in node (edge) characteristics at any time, whereas discrete dynamic graphs allow changes only to occur in timeslices.  DynNoSlice implements the above model as a collection of piecewise, linear functions defined on intervals in the space-time cube. Fast access to these functions at any given time is required. In our implementation, we use interval trees.  When the intervals are guaranteed to be non-overlapping, simpler structures such as binary search trees can be used.
%However, timeslices can be computed at any time $t$ creating a static graph $G_t$ with the nodes and edges marked as present in $t$ by the appearance functions.
%
%Thus, the complexity of the function representing a node in the space-time cube can adjust %its complexity 
%independently based on its interaction with other graph elements over time.  In dynamic graphs drawn with timelices this complexity is imposed by the timeslicing.  %Moreover, after the piecewise linear functions are drawn in the space-time cube,

\section{DynNoSlice Implementation}
A continuous dynamic graph $D$ can be transformed into a 3D static graph $D'$ by embedding it in the space-time cube. Algorithms for static or discrete dynamic graphs can be extended to work with this new representation. In this section, we describe our force-directed algorithm for drawing continuous dynamic graphs.  It has been implemented and the source code is available\footnote{\url {http://cs.swan.ac.uk/~dynnoslice/software.html}}.

\subsection{Representation in the Space-Time Cube}
We can define a space-time cube transformation (STCT) that transforms a continuous dynamic graph into a drawing in the space-time cube. In $D'$, the presence and position of each node is represented by a sequence of trajectories.  The shapes of these trajectories are defined by the position attributes. In the example above, the trajectory of $v$ is defined by three line segments, $(1,0,9) \rightarrow (4,0,12)$, $(2,-2,12) \rightarrow (5,1,15)$ and $(5,1,17) \rightarrow (4,5,19)$, and by portions of the line $(0,0,x)$ (see Fig.~\ref{FIG:Transformation_1}). The number of these trajectories is also affected by the appearance attribute. In the example above, the node $v$ appears two times: at $[2,7)$ and at $(9, 13]$. Clearly, the behavior of the node at times when it is not part of the graph is non-influential. Therefore, the node appearance and position in the space-time cube can be identified by the segments $s_{v,1} = (0,0,2) \rightarrow (0,0,7)$, $s_{v,2} = (1,0,9) \rightarrow (4,0,12)$ and $s_{v,3} = (2,-2,12) \rightarrow (3,-1,13)$  (see Fig.~\ref{FIG:Transformation_2}). The trajectory given by the polyline is made of the start and end points, as well as the bends (the junctions of consecutive segments).

The representation of edges in the space-time cube is less intuitive. By connecting two trajectories with lines in the space-time cube, we obtain a ruled surface. Therefore, an edge $e=(u,v)$ is a surface that connects trajectories $u$ and $v$ for a duration indicated by $\mathcal{A}_e$. If the node trajectories are not continuous as in the above example:
\[ \lim_{t \rightarrow 12^{-}} \mathcal{P}_v(t) = (4,0) 
\qquad \text {and} \qquad
\lim_{t \rightarrow 12^{+}} \mathcal{P}_v(t) = (2,-2), \]
an edge might create two or more surfaces  (see Fig.~\ref{FIG:Transformation_3}).

If the trajectory segments, considered as vectors, form an acute angle with respect to the positive time axis, this transformation is easily invertible. Thus, trajectory segments cannot fold back on themselves, as it would identify multiple positions for that node at a given time. We can then work on this continuous dynamic graph in 3D as if it were a static graph as follows:
\[ D_a \ \rightarrow \ \text{STCT} \ \rightarrow \ D_a' \ \rightarrow \ \text{\emph{Operation}} \ \rightarrow \ D_b' \ \rightarrow \ \text{STCT}^{-1} \ \rightarrow \  D_b \]

%In order to draw the continuous dynamic graph in the space-time cube, we need to convert time units to space units.  The appendix of this paper provides further detail.

\subsection{Force-Directed Drawing Algorithm}
Our force-directed algorithm is based on earlier variants by Simonetto \ea~\cite{ImPrEd,EulerSmooth}. As in most force-directed algorithms, after an initialization phase, the algorithm iteratively improves the current layout of the drawing in 3D for a given number of iterations in the following way:
\begin {itemize}
\item For each iteration of the algorithm:
\begin {itemize}
\item Compute and sum the forces based on the force system.
\item Move nodes based on these forces and the constraints.
\item Adjust trajectory complexity in the space-time cube.
\end {itemize}
\end {itemize}

\subsubsection{Initialization}
Each node $v$ is randomly assigned a position $(a,b)$ in the 2D plane. %Being the random 2D position of the node $v$, t
The points defining $v$'s trajectory are extruded linearly along the time axis $(a,b,x)$, where $x$ is the time coordinate.% calculated as mentioned above.

\subsubsection{Forces}
DynNoSlice has five forces.  The first three forces adapt standard, force-directed approaches to work with trajectories in the space-time cube. The final two are novel for continuous dynamic graph drawing. In our notation, a star transforms 3D vector to 2D by dropping the time coordinate. For example, if $p=(1, 2, 3)$ then $p^*$ is $(1, 2)$. The parameter $\delta$ is an ideal distance between two node trajectories. %which is the distance at which the attraction and repulsion forces balance each other. 

{\it 1. Node Repulsion.}  This force repels trajectories from each other. % when they are too close. 
The force evenly distributes node trajectories in space and prevents crowding~\cite{88Pylyshyn,05Liu}.

\begin{figure}[t]
    \centering
\subfloat[]{\includegraphics[width=0.17\linewidth]{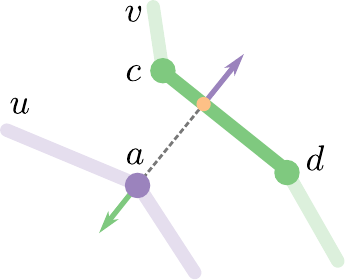}\label{FIG:NodeRepulsion_1}} \hspace{.2cm}
\subfloat[]{\includegraphics[width=0.17\linewidth]{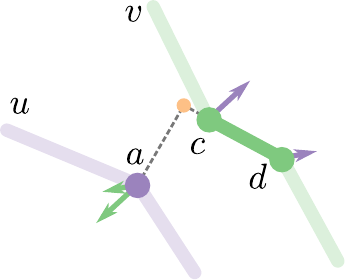}\label{FIG:NodeRepulsion_2}}
\hspace{.2cm}
\subfloat[]{\includegraphics[width=0.17\linewidth]{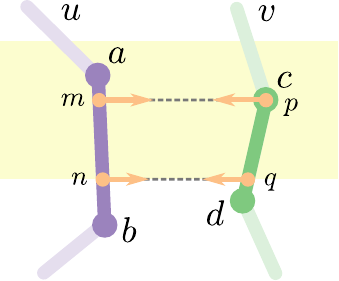}\label{FIG:EdgeAttraction_1}} \hspace{.2cm}
\subfloat[]{\includegraphics[width=0.18\linewidth]{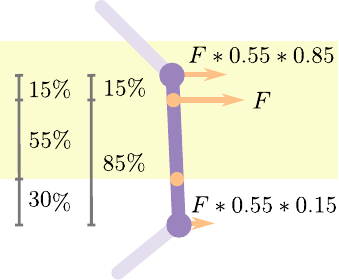}\label{FIG:EdgeAttraction_2}}
\hspace{.2cm}
\subfloat[]{\includegraphics[width=.12\linewidth]{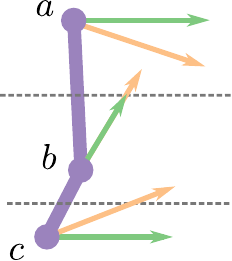}\label{FIG:timeconstraint}}
\medskip\caption{\small Forces and constraints.  \sflabel{a} and \sflabel{b} Node repulsion force between point $a$ and the line segment $c \rightarrow d$ that represents the trajectories of nodes $u$ and $v$.  \sflabel {c} and \sflabel{d} Edge attraction for an edge in the interval highlighted with yellow background.  \sflabel{e} Time movement restriction. Endpoints ($a$ and $c$) must keep their assigned time coordinates.  Bend $b$ cannot move past half the distance with other bends or endpoints.} %\sflabel{a} When the points $a$, $c$ and $d$ are not aligned, $a$ is projected into the plane defined by $c \rightarrow d$. In this case, EdgeNodeRepulsion($\delta$) computes a force that repels the segment and the point. \sflabel{b} If the projection is not on the segment, EdgeNodeRepulsion($\delta$) computes repulsive forces between the point and the segment endpoints. We apply this case even when the points $a$, $c$ and $d$ are collinear and it is not possible to identify a projection.}
\label{FIG:NodeRepulsion}
\end{figure}

For each segment endpoint $a$ in the trajectory of node $u$ and segment $s_{v,j} = c \rightarrow d$ of node $v$, with $u \neq v$, we compute the forces generated on the points $a$, $c$ and $d$ (see Fig.~\ref{FIG:NodeRepulsion}). If the points $a$, $c$ and $d$ are not collinear and ($p \in s_{v,j}$), they form a plane. In this case, we apply the force EdgeNodeRepulsion($\delta$) described in previous work~\cite{EulerSmooth}, except node positions are in 3D. If the points are collinear or the projection $p$ of $a$ does not fall in the segment $s_{v,j}$ ($p \notin s_{v,j}$), we apply  NodeNodeRepulsion($\delta$)~\cite{EulerSmooth} between $a$ and $c$ and between $a$ and $d$.

%For each segment endpoint $a$ in the trajectory of node $u$ and segment $s_{v,j} = c \rightarrow d$ of node $v$, with $u \neq v$, we compute the forces generated on the points $a$, $c$ and $d$ (see Fig.~\ref{FIG:NodeRepulsion}). If the points $a$, $c$ and $d$ are not collinear, they form a plane. In this case, we apply the force EdgeNodeRepulsion($\delta$) described in previous work~\cite{EulerSmooth}, except node positions are in 3D. If the points are collinear, we apply  NodeNodeRepulsion($\delta$)~\cite{EulerSmooth} between $a$ and $c$ and between $a$ and $d$ which is consistent with the case in which the points are not collinear and the projection $p$ of $a$ does not fall in the segment $s_{v,j}$ ($p \notin s_{v,j}$). 

Since distant segments do not interact significantly, they can be ignored to reduce the running time. A multi-level interval tree is used to identify segments that are sufficiently close ($< 5\delta$).  All other pairs are ignored.

{\it 2. Edge Attraction.} This attractive force pulls trajectories that are linked by an edge closer to each other. The force is exerted only for the intervals where the edge is present. %allowing the trajectories of the related nodes to not influence each other outside of these time intervals.
%
%\begin{figure}[t]
%\centering
%\subfloat[]{\includegraphics[width=0.18\linewidth]{EdgeAttraction_1}\label{FIG:EdgeAttraction_1}} \hspace{.5cm}
%\subfloat[]{\includegraphics[width=0.18\linewidth]{EdgeAttraction_2}\label{FIG:EdgeAttraction_2}}
%\caption{Edge attraction for an edge between nodes $u$ and $v$ in the interval highlighted with yellow background.} %\sflabel{a} We compute the points $m$ and $p$ as those at the highest level inside both segments and inside the edge presence interval and points $n$ and $q$ as those at the lowest level. An attraction force is computed between $m$ and $p$ and between $n$ and $q$ using the EdgeContraction($\delta$) formula. \sflabel{b} Each force of an edge is distributed to its endpoints according to the coverage of the segment over the edge interval and the closeness of the force to the segment endpoint.}
%\label{FIG:EdgeAttraction}
%\end{figure}

Let us consider an edge $e = (u,v)$ that appears at interval $(t_1, t_2)$ and let $I = (t_1 \tau, t_2 \tau)$ be the transformed time interval in the space-time cube with conversion factor $\tau$ that transforms time into the third dimension of the space time cube. For each pair of segments $s_{u,i} =  a \rightarrow b$ and $s_{v,j} = c \rightarrow d$ that overlap with the interval $(f,g)$, we compute the points $m,n$ of segment $s_{u,i}$ and $p,q$ of segment $s_{v,j}$ so that:
\[ \min \big\{f \in I : \exists m \in s_{u,i}, \ \exists p \in s_{v,j}, \ f = m[2] = p[2]\big\} \]
\[ \max \big\{g \in I : \exists n \in s_{u,i}, \ \exists q \in s_{v,j}, \ g = n[2] = q[2]\big\} \]
where $x[2]$ is the time coordinate of the point $x$ in the space-time cube. We compute the attractive force between the points $m$ and $p$, and $n$ and $q$, using EdgeContraction($\delta$)~\cite{EulerSmooth} (see Fig.~\ref{FIG:EdgeAttraction_1}). This force is applied to the segment endpoints once scaled by its distance from the application point and by the coverage of the edge appearance on the segment (see Fig.~\ref{FIG:EdgeAttraction_2}). For example, if the force $F_m$ attracts $m$ to $p$, then $F_a$ applied to $a$ will be:
\[ F_a = F_m * \frac{a[2] - m[2]}{a[2] - b[2]} *  \frac{n[2] - m[2]}{a[2] - b[2]}. \]

{\it 3. Gravity.} This force encourages a compact drawing of node trajectories. Let $c$ be the center in 2D of the initial node placement in the space-time cube. The gravity $F$ of each segment endpoint $a$ is $F = c^* - a^*$.

{\it 4. Trajectory Straightening.} This force smooths node trajectories, helping with node movements over time. For trajectory bends, we use the CurveSmoothing~\cite{EulerSmooth} force, which pulls a bend $b$ to the centroid of the triangle $\Delta abc$ formed by using the previous and next bends or endpoints $a$ and $c$. A trajectory endpoint $a$ has no such triangle.  Therefore, it is pulled in 2D toward the midpoint of the segment formed with the closest bend or endpoint $b$.

{\it 5. Mental Map Preservation.} This force prevents trajectory segments from making large angles with respect
to time. When segments form a \ang{90} angle with time, a node essentially ``teleports'' from one place to another, while segments parallel to the time axis result in no node movement. Thus, segments should form small angles with the time axis. This force pulls endpoints $a$ and $b$ of each trajectory towards each other in 2D with a magnitude based on the angle $\alpha$ the segment makes with the time axis:
\[  F_a = (b^*-a^*) * \frac{\alpha}{\ang{90} - \alpha} \]

\subsubsection{Constraints}
Node movement constraints ensure valid drawing of the continuous dynamic graph in the space-time cube. In particular, constraints are needed to prevent undesired movements.

{\it Decreasing Max Movement.} We insert a constraint on the maximum node movement allowed at each iteration. This allows large movements at the beginning of the computation, and smaller refinements towards the end. This constraint is similar to DecreasingMaxMovement($\delta$)~\cite{EulerSmooth}.

{\it Movement Acceleration.} This constraint promotes consistent movements with previous iterations and penalizes movements in the opposite direction. This constraint corresponds to MovementAcceleration($\delta$)~\cite{EulerSmooth}.

{\it Time Correctness.} This constraint prevents a node from changing its time coordinate. Consider the trajectory formed by the segment $t = a \rightarrow b$. If $a$ changes its time coordinate in the space-time cube, the time of its appearance will also change. Now, consider a trajectory formed by several segments, $t = a \rightarrow b \rightarrow c \rightarrow d$. The trajectory endpoints $a$ and $d$, corresponding to the appearance and disappearance of the node, should have fixed time positions.  However, bends $b$ and $c$ can move in time, so long as they do not pass each other ($a[2] < b[2] < c[2] < d[2]$) as this would result in a node being in two locations at once.
Therefore, the movement $M_a$ of node $a$ in time is constrained to be:
$M_a \leftarrow M_a^*$
if $a$ is the endpoint of a trajectory, and 
\[ M_b \leftarrow M_b * \max \left\{r \in [0,1] : \frac{M_a[2] - M_b[2]}{2} < r M_b[2] < \frac{M_c[2] - M_b[2]}{2} \right\} \]
if $b$ is a trajectory bend between other bends or endpoints $a$ and $c$ (see Fig.~\ref {FIG:timeconstraint}).

\subsubsection {Complexity Adjustment of Node Trajectories}
As bends move freely in time and space, node trajectories can be oversampled or undersampled. Therefore, bends can be inserted or removed from the polyline representing a node trajectory.  
%This concept is derived from the \emph{flexible edges} feature~\cite{ImPrEd,EulerSmooth} used in Euler diagram visualisation methods. 
If a segment of the trajectory between bends $a$ and $b$ is greater than a threshold ($2\delta$ in this paper), a bend is inserted at its midpoint. Similarly, if two consecutive segments $ab$ and $bc$ are placed such that the distance between $a$ and $c$ is less than a threshold ($1.5\delta$ in this paper), the bend $b$ is removed and the two segments are replaced by $ac$. 

%\begin{figure}[t] \centering
%\begin{minipage}[t]{0.2\linewidth}
%\includegraphics[width=\linewidth]{Movement_restriction}
%\caption{Time movement restriction for trajectories. Endpoints ($a$ and $c$) must keep their assigned time positions. Bends ($b$) cannot move past half the distance with other bends or endpoints.} \label{FIG:Movement_restriction}
%\end{minipage}
%\hspace{1cm}
%\begin{minipage}[t]{0.3\linewidth}
%\includegraphics[width=\linewidth]{Time_flattening}
%\caption{Node (edge) transparency when flattening the time dimension. We consider a Gaussian curve centered at $t$ and with standard deviation equal to a sixth of the cluster interval in time. The appearance of the graph element in time is extracted (green lines). The transparency of the element is the area under the Gaussian during the appearance intervals (yellow) divided by the total area under the curve.} \label{FIG:Time_flattening}
%\end{minipage}
%\end{figure}

\section{Evaluation}
We perform a metric-based evaluation of our approach against the state-of-the-art algorithm \visone~\cite{Visone} to demonstrate the advantages of our model.% for drawing continuous dynamic graphs.

\subsection{Data Sets}
\label {secData}
%We use data sets that are continuous in time and data sets that have timeslices.  In the appendix, we provide extra information on how we process these data sets.  

{\it InfoVis Co-Authorship (Discrete):}  a co-authorship network for papers published in the InfoVis conference from 1995 to 2015~\cite{CiteVis_data}. Authors collaborating on a paper are connected in a clique at the time of publication of the paper. Note this is not a cumulative network as authors can appear, disappear, and appear again.  The data is of discrete nature with exactly 21 timeslices (one per year). 

{\it Van De Bunt (Discrete):} shows the relationships between 32 freshmen at seven different time points.
%At each point in time, each participant was asked to rate the friendship into one of several categories spanning ``best friendship'' to ``troubled relation''. Additional attributes such as gender, smoker/non-smoker, and program duration are also provided.
A discrete dynamic graph is built using the method of Brandes \ea~\cite{Stress_comparison}, with an undirected edge inserted into a timeslice if the participants reciprocally report  ``best friendship'' or ``friendship'' at that time.

{\it Newcomb Fraternity Data (Discrete):}  contains the sociometric preference of 17 members of a fraternity in the University of Michigan in the fall of 1956~\cite{Newcomb}.
%Each participant was asked to order the other students from their closest to their least close friend. This data was collected weekly for about 15 weeks.
As in previous work~\cite{Stress_comparison}, at each timeslice, we inserted undirected edges connecting students to their three best friends.

{\it Rugby (Continuous):} is a network derived from over 3,000 tweets involving teams in the Guinness Pro12 rugby competition.  The tweets were posted between 09.01.2014 and 10.23.2015. Each tweet contains information about the involved teams and the time of publication with a precision down to the second.

{\it Pride and Prejudice (Continuous):} lists the dialogues between characters in the novel {\it Pride and Prejudice} in order~\cite {16Grayson}. The book has 61 chapters and the data set includes over 4,000 interactions between characters. 

\subsection{Method} \label{secMethod}
As there are no continuous dynamic graph drawing algorithms, we need to compare our results with discrete dynamic graph drawing algorithms.  In our metric evaluation, we considered three drawing approaches:
\begin{itemize}
\item Visone (v) drawings were computed using the discrete version (timesliced) with \visone~\cite{Visone}. %\footnote{\url{http://visone.info}}
  We use a linking strategy~\cite{Stress_comparison} with a default link length of 200 and a stability parameter $\alpha=0.5$. %current layout quality. 
 \item Discrete (d) drawings were computed using a modified version of DynNoSlice on the discrete version.  Each trajectory bend coincides with a timeslice and bends cannot be inserted or removed. The rest of the algorithm is unaltered.  We used $\delta=1$ as the desired edge length parameter.
 \item Continuous(c) drawings were computed using DynNoSlice. We used $\delta=1$ as the desired edge length parameter.
\end{itemize}

We evaluate the results using a variety of metrics:
\begin{itemize}
 \item Stress: the average edge stress, as defined by Brandes and Mader~\cite[Formula 1]{Stress_comparison}. As stress is defined for a static graph, we slice the space-time cube and average the stress computed on each timeslice.
 \item Node Movement: the average 2D movements of the nodes. Intuitively, this is the average distance traveled by nodes when animating the dynamic graph.
 \item Crowding: This metric counts the number of times nodes collide during the animation of the dynamic graph. 
 \item Running Time in seconds. %: the time required to produce a drawing.
\end{itemize}

For continuous dynamic graphs, we can compute the stress of the nodes and edges present in the timeslice defined by the discrete dynamic graph or on the node and edge set at that precise point in continuous time. We can also compute stress between timeslices.
%Thus, we have two node and edge set definitions for stress. Also, we compute stress on the timeslices defined by the discrete dynamic graph and \textit {between} timeslices by selecting a sampling frequency ten times that of the discrete sampling rate.
Thus, we consider four measures of stress:
\begin{itemize}
 \item StressOn(d): the stress computed on the timeslices using the node and edge set of that timeslice.
 \item StressOff(d): the stress computed on and between the default timeslices using the node and edge set of the closest timeslice in time, when between two timeslices.
 \item StressOn(c): the stress computed on the timeslices using the precise node and edge appearances in continuous time.
 \item StressOff(c): the stress computed on and between the timeslices using the precise node and edge appearances in continuous time.
\end{itemize}

\subsubsection{Graph Scaling}
Uniformly scaling node positions changes the measure of stress even though the layout is the same~\cite {gansner2013maxent,kobourov2014crossings,ortmann2016sparse}. In order to compare methods as fairly as possible, we used a strategy where scale-independent values of stress are compared as follows. Both \visone and DynNoSlice have a parameter that indicates the desired edge length. First, we verified that \visone  produces the same result (up to scale) when changing the edge length parameter.  
%with different scaling when choosing different edge length parameters.
% Indeed, by choosing an edge length of 400 instead that 200 we obtained a layout where nodes have the same position but that is twice as large. 
%We can therefore use any edge length of our choice (we kept the default value of 200),
Thus, we use the default value of edge length but consider different scaling factors to compare to the output of our algorithm.
%the drawing to the proportions used in our algorithm.  
For the experiment, we defined nodes to be circles of diameter 0.2 and with an ideal edge length of 1. To obtain such drawing with our approach, we run the algorithm with an ideal edge length parameter $\delta=1$. To obtain such drawing with \visone, we run it with the default edge length of 200 and scale it down by a factor 200.  %These scaled network series are passed to the next stage of our procedure.

Related work in static graph drawing~\cite {gansner2013maxent,kobourov2014crossings,ortmann2016sparse} searches for the best scaling factor via binary search, as a minimum is guaranteed.  For our metric (average stress across all timeslices) we have no such guarantee.  Thus, we evaluate scaling factors $(1.1)^i$, with $i \in \Z : -20 < i < 20$ for the best  StressOn(d) value. This scaling factor is used to compute all metrics.  After plotting the average stress for each data set, a minimum in this range was consistently observed.%, which are reported in Table~\ref{TAB:metrics}.

%\DTLloaddb{dataTable}{data/data.csv}
%\begin{table*}[t]
%\renewcommand*{\dtlrealformat}[1]{\DTLround{\rounded}{#1}{2}\rounded}
%\DTLdisplaydb{dataTable}
%\caption{Data} \label{TAB:metrics}
%\end{table*}

\subsection {Results}
Videos of {\it Newcomb}, {\it Rugby}, and {\it Pride and Prejudice} are provided in the supplementary material.  In this section, we present our quantitative results.

\visone is often faster than our continuous approach as DynNoSlice operates on a greater volume of data -- the data between timeslices.  The discrete version of DynNoSlice is often slower than both, as the continuous approach is more naturally expressed without timeslices.  Therefore, there is a penalty for imposing timeslices on it. 

Table~\ref{tableDiscrete} shows the results on the discrete data sets.  Continuous stress metrics are not computed, because nodes and edges only appear on timeslices.  In the VanDeBunt and Newcomb data sets, \visone %has the lowest scaling factor and 
outperforms DynNoSlice in terms of stress.  Movement and crowding are comparable among all three approaches.  Our continuous approach is sometimes able to improve on \visone when stress is computed off timeslices.  When comparing the discrete and continuous versions of DynNoSlice, our discrete version is often able to optimize on-timeslice stress. InfoVis is an outlier for the timesliced data sets.  In particular, our continuous approach outperforms \visone in terms of stress.

Table~\ref {tableContinous} shows the results of our metric experiment on the continuous data sets. Our continuous approach %consistently 
has lower off-timeslice stress, lower average movement, fewer crowding events, and occasionally lower on-timeslice stress.  
%Also, the approach has consistently lower average movement and fewer crowding events.

\begin {table}[t]
  \centering
  {\scriptsize
\begin {tabular} {rcSSSSSS}
\toprule
\textbf {Graph} & \textbf {Type} & \textbf {Time (s)} & \textbf {Scale} & \textbf {StressOn (d)} & \textbf {StressOff (d)} & \textbf {Movement} & \textbf {Crowding}\\
\midrule
VanDeBunt &v&0.13&1.00&1.14&1.46&3.80&0\\
 &d&7.73&0.62&1.20&1.20&3.91&0\\
 &c&6.73&0.68&1.19&1.29&3.69&0\\
\midrule
Newcomb &v&0.11&1.00&14.04&14.77&16.36&8\\
 &d&9.68&0.68&16.61&16.59&13.48&2\\
 &c&7.62&0.75&18.13&17.98&12.37&0\\
\midrule
InfoVis  &v&77.43&0.47&51.66&52.98&2.15&36\\
 &d&388.38&0.56&31.09&31.04&2.03&8\\
 &c&381.15&0.56&32.70&34.02&1.91&6\\
\bottomrule
\end {tabular}
}
\medskip\caption {\small Results for the discrete data sets on our metrics.}  %All metrics are as defined in Section~\ref{secMethod}.  As the data is timesliced, only discrete edge sets can be measured as continuous time information is missing.}
\label {tableDiscrete}
\end {table}

\begin {table}[h]
  \centering
  {\scriptsize
\begin {tabular} {rcSSSSSS}
\toprule
\textbf {Graph} & \textbf {Type} & \textbf {Time (s)} & \textbf {Scale} & \textbf {StressOn (c)} & \textbf {StressOff (c)} & \textbf {Movement} & \textbf {Crowding}\\
\midrule
Rugby &v&0.08&0.68&3.08&2.71&25.47&6\\
 &d&7.40&0.68&1.84&1.71&16.23&1\\
 &c&3.88&0.51&1.84&1.77&6.57&0\\
\midrule
Pride  &v&3.39&0.18&0.62&0.88&5.44&682\\
  &d&1655.50&0.32&0.82&0.86&6.95&13\\
  &c&75.61&0.24&0.83&0.85&1.12&3\\
\bottomrule
\end {tabular}
}
\medskip\caption {\small Results for the continuous data sets on our metrics.}%  All metrics are as defined in Section~\ref{secMethod}.  As the data is continuous, we measure stress on the continuous set of edges to make the algorithms comparable.}
\label {tableContinous}
\end {table}

\subsection {Discussion}

Our results on the discrete data sets were expected:  \visone optimizes for stress directly on every timeslice and so outperforms our force system in terms of stress, while it is comparable in terms of movement and crowding.  As a state-of-the-art algorithm for timesliced graph drawing, it is difficult to compete with \visone when it is running on the type of data for which it was designed.  However, when stress is measured off the timeslices, our continuous approach often outperforms \visone.  The timesliced model does not allow for stress to be optimized between timeslices and must resort to linear interpolation, leading to suboptimal stress.  In our continuous dynamic graph model, we optimize for stress in continuous time, leading to this performance improvement.
The InfoVis data set is an exception where DynNoSlice is able to improve on \visone in terms of stress.  
%At first glance this may seem strange as \visone is optimising stress on the timeslices directly.  However, t
This result could be due to the bursty nature of this graph (edges are only present if two authors published a joint paper that year).  Therefore, large parts of the graph change drastically from year to year.  Allowing node trajectories to evolve independently of timeslices may allow DynNoSlice to perform better.% improve on \visone in this case.  
%These drastic changes in graph structure causes the data set to behave more like a continuous one.  

%On the continuous data sets, our results are consistently very different.  Nearly 
For nearly all continuous data sets, our continuous model often outperforms \visone in terms of stress, movement, and crowding.  This is likely due to the fact we do not use timeslices to compute the layout of the dynamic graph.  As a result, we are able to optimize stress between timeslices as well.  On-timeslice stress is an exception, as it is directly optimized by \visone.

It is surprising that our continuous dynamic graph drawing algorithm simultaneously improves node movement and crowding while remaining competitive or improving on stress. %when compared to \visone on  
This finding may seem counter-intuitive as low stress usually corresponds to high node movement.  This result can be explained by the fact that nodes in our continuous models are polylines of adaptive complexity in the space-time cube.  Nodes with few interactions in the data will be long straight lines, potentially passing through many timeslices.  These areas of low complexity will reduce average node movement.  In a model that uses timeslices, each timeslice is forced to have that node with inter-timeslice edges.  Therefore, timeslices impose additional node movement that may not be necessary.

In terms of crowding, the continuous model allows the polyline representing a node to adapt its complexity between timelices if there are many interactions.  When there are many changes to the graph in a short period of time, these polylines have increased complexity, allowing nodes to avoid crowding.  In a timesliced model, only linear interpolation is possible, and all nodes must follow straight lines. Thus, crowding is incurred.  Crowding is also avoided in our continuous model as our polylines have repulsive forces between them.  In all timeslice approaches, inter-timeslice edges do not repel each other, potentially causing crowding events.

\section {Conclusions and Future Work}
%In this paper, 
We presented a model for dynamic graph drawing without timeslices 
%where timeslices do not exist.  
%In this model, 
in which 
nodes and edges, along with their attributes, are defined on continuous time intervals.  We developed a dynamic graph drawing algorithm, DynNoSlice, that visualizes graphs in this model by working on the space-time cube.  An implementation of this algorithm is available along with a video. %showing its output.
%Although a natural way to visualize continuous graphs is an animation, we also develop a method for selecting interesting timeslices in order to support a small multiples visualization of these data sets.
In our evaluation, we demonstrate that our continuous approach has significant advantages over timeslicing the continuous data.

The focus of this paper is a method to draw dynamic graphs without timeslices. An animation of a slice traversing the layout in the space-time cube is a natural visualization.  However, animation is not always effective in terms of human performance~\cite {02Tversky,10FarrugiaIVJ,10ArchambaultTVCG}, especially when events are of short duration.  More effective visual representations the layout present in the space-time cube are necessary.

The primary issue that we have found with collapsing continuous time down onto a series of timeslices is that the timeslices could oversample/undersample the data in the continuous dimension.  In signal processing, the Nyquist frequency gives the minimum sampling rate required to reconstruct the signal.  Regular timeslices taken a the smallest temporal distance between two events should be sufficient to avoid undersampling, but lower sampling frequencies could be possible and remain future work. 

\newpage
\bibliographystyle{abbrv-doi}

\bibliography{Bibliography}
\end{document}